\documentstyle[12pt,aasms4]{article}
\begin{document}
\title{Zeroing the Stellar Isochrone Scale: The Red Giant Clump Luminosity at Intermediate
Metallicity}
\author{Bruce A. Twarog, Barbara J. Anthony-Twarog, and Andrew R. Bricker}
\affil{Department of Physics and Astronomy, University of Kansas,
Lawrence, KS  66045-2151; twarog@kustar.phsx.ukans.edu, anthony@kubarb.phsx.ukans
.edu}

\begin{abstract}
The color-magnitude diagrams of the open clusters NGC 2420 and NGC 2506 
have been investigated as intermediate links between the solar neighborhood
and the Magellanic Clouds. Two sets of theoretical isochrones which
include convective overshoot are zeroed to the sun
at solar abundance and to the unevolved main sequence dwarfs of the
{\it Hipparcos} catalog at [Fe/H] = --0.4, requiring a differential of
0.4 mag between the unevolved main sequences at a given color. Adopting
$E(B-V)$ = 0.04 and [Fe/H] = --0.39 for NGC 2506 and $E(B-V)$ = 0.04 and
[Fe/H] = --0.29 for NGC 2420, the respective apparent moduli are $(m-M)$ = 
12.70
and 12.15, while the ages of both clusters are approximately 1.9 $\pm$ 0.2 Gyr
or 2.2 $\pm$ 0.2 Gyr, depending on the choice of isochrones. 
From the composite giant branch of the two clusters, the mean
clump magnitudes in $V$ and $I$ are found to be 
+0.47 and --0.48 (--0.17,+0.14), respectively.
Applying a metallicity correction to the $M_I$ values,
the cluster sample of Udalski (1998)\markcite{u1} leads to $(m-M)_0$ = 18.42 
(+0.17,--0.15)
and 18.91 (+0.18,--0.16) for the LMC and SMC, respectively. A caveat to
this discussion and potentially to the claim that
clusters of the same abundance and age are identical is the observation
that the $(V-I)$ colors of the red giants in NGC 2506 are significantly redder
at a given $(B-V)$ than the giants in clusters of comparable age and/or
metallicity. The distance scale above has been derived using the general
cluster relation between $(B-V)$ and $(V-I)$. If the CCD photometry in
NGC 2506 is correctly tied to the standard system, $M_I$ for the clump will
decrease and the distance moduli should increase by 0.1 mag.

\end{abstract}
 
\section{Introduction}
Within the quest to define the distance scale of the nearby Universe, the
Magellanic Clouds have played a critical role as an intermediate link
between local distance indicators and those needed to reach galaxies
beyond the Local Group. As summarized in Fig. 1 of 
Cole (1998)\markcite{c3}, virtually
every potential means of fixing the location of this datum within the
grand scale has been attempted in recent years, many of the approaches 
catalyzed by the availability of precise parallaxes from {\it Hipparcos}.
A recent technique which has generated both discussion and controversy is 
the location of the red giant clump (RGC), the locus of stars in the 
post He-flash phase of evolution, generally analogous to the horizontal 
branch in globular
clusters. The rationale behind this choice is obvious; the clump stars are
intrinsically bright, well separated from the redder, first-ascent giants,
and the luminosities appear to be only weakly dependent on age and metallicity,
particularly in the infrared. Unfortunately, applications of the approach
have produced significantly different results. Cole (1998)\markcite{c3} 
has derived a
true distance modulus of $(m-M)_0$ = 18.36 $\pm$ 0.17, consistent within 
the errors with the Cepheid
Key Project value of 18.50 $\pm$ 0.10, while 
Stanek et al. (1998)\markcite{s5} find 18.06
$\pm$ 0.03 $\pm$ 0.09. (Note, throughout the text the apparent modulus, i.e.,
uncorrected for reddening, is designated by $(m-M)$ without the subscript 0.)
The primary source of disagreement lies with the 
metallicity and age sensitivity of the clump luminosity. The 
absolute $I$ magnitude of
the clump is based upon the {\it Hipparcos} (Perryman et al. 1997)\markcite{p2}
results for a large sample
of nearby giants (Stanek \& Garnavich 1998\markcite{s4}). Though 
it has long been known
that the LMC is more metal-deficient than the solar neighborhood, a result
confirmed by the color distribution of the clump stars (see, e.g., Fig.
2
of Stanek et al. 1998\markcite{s5}), considerable empirical evidence 
exists that these
metallicity differences should not significantly alter the luminosity of
the clump (Paczy\'{n}ski \& Stanek 1998\markcite{p1}; Stanek \& Garnavich 
1998\markcite{s4}; Udalski et al.
1998\markcite{u2}; Stanek et al. 1998\markcite{s5}; 
Udalski 1998)\markcite{u1}. This empirical evidence is 
challenged by Cole
(1998)\markcite{c3} who uses theoretical models of clump stars to demonstrate that
mass and metallicity can significantly impact the luminosity distribution 
of clump stars. If the theoretical models are reliable representations
of He-burning stars, the mean luminosity of the RGC will depend upon the star
formation and chemical history of the population under discussion, i.e.,
it isn't necessarily appropriate to apply the solar neighborhood results 
to the LMC (see, e.g., Beaulieu \& Sackett 1998)\markcite{b1}.

Ideally, one could minimize, if not resolve, this issue if a population
of metal-deficient stars with ages typical of the LMC could be identified
locally and used in the comparison. While the field population 
of {\it Hipparcos}
appears to contain a modest sample of clump giants with [Fe/H] near --0.4 (see
Fig. 5 of Jimenez et al. 1998\markcite{j2}), there is
at present no means of dating these stars. If they are typical of the 
old/thick disk, their ages may be closer to 10 $\pm$ 2 Gyr (Wyse \& Gilmore
1995)\markcite{w1} than the more
common 2 $\pm$ 2 Gyr found at this abundance within the LMC. Moreover,
the abundances derived for the clump stars by Jimenez et al.
(1998)\markcite{j2} using DDO photometry and the calibration of Janes
(1975\markcite{j3}, 1979\markcite{j4}) are systematically in error.
As discussed in detail in Twarog et al. (1997)\markcite{t5}, the revised
DDO calibrations by both Piatti et al. (1993)\markcite{p4} and Twarog
\& Anthony-Twarog (1996)\markcite{t6} demonstrate that the original
[Fe/H] calibration systematically underestimates the red giant abundances
by between 0.1 and 0.2 dex. On the cluster abundance scale adopted in 
this investigation, giants with [Fe/H] = --0.5 from H$\o$g \& Flynn
(1998)\markcite{h2} have actual abundances near --0.3.
 
Somewhat overlooked
in the discussion of this problem is the fact that a sample of stars with
well-defined ages and compositions comparable to those found in the LMC
does exist in the form of the open clusters of the galactic anticenter.
These clusters have typical [Fe/H] between --0.2 and --0.5 (Twarog et al. 
1997)\markcite{t5}
and ages between 1 and 4 Gyr (Friel 1995)\markcite{f1}. The primary 
weakness in the
use of these objects has been the uncertainty in the distance determination, 
based upon
main sequence fits to theoretical isochrones assumed to have similar 
compositions and/or differential comparisons to other clusters via the
main sequence or the RGC. In the end, the uncertainties in
the absolute and differential scales of the isochrones translate into
comparable uncertainties in the absolute luminosities of key color-magnitude
diagram (CMD) features such as the clump. With the availability of the
{\it Hipparcos} data, these difficulties can be resolved through the use
of main sequence stars rather than clump giants, i.e., we can link the LMC
to the solar neighborhood via the anticenter clusters.

The goals of this investigation are twofold:
First, we will attempt to zero correctly the theoretical
isochrones which are commonly used in age and distance determination via
direct comparison with nearby field stars of comparable metallicity. This
will be the focus of Sec. 2.

Second, using the observational data for two virtually identical clusters,
NGC 2420 and NGC 2506, as detailed in Sec. 3, we will derive the distance
moduli and cluster ages through comparison with the appropriate isochrones
and, indirectly, fix the absolute magnitude of the RGC
for a 2 Gyr old population with
[Fe/H] of  --0.4.  Sec. 4 provides the straightforward
comparison to the LMC and SMC and a discussion of our results.

\section{Zeroing the Scale}
A critical step in the use of any set of theoretical isochrones 
is the transition
between the theoretical [$M_{bol}$, log $T_e$] and the 
observational [$M_V$, $(B-V)$] plane. The
required transformations are a function of composition and, to a lesser degree,
surface gravity; they can be derived via empirical relations and/or model
atmospheres convolved with appropriate filter functions. Once transformed,
the observational relations can be tested by comparison to 
well-defined sequences found
in nearby clusters, as illustrated in VandenBerg (1985)\markcite{v1} 
and VandenBerg \&
Poll (1989)\markcite{v2}. Though such comparisons permit adjustments 
which guarantee that
the slopes of the theoretical color-magnitude relations will resemble
real clusters, they resolve only half the problem in that the zero point
of the scale is still unknown. By the zero point we refer to the link to
the two additional fundamental parameters which uniquely define a star's
structure: mass and age. (Note that age is equivalent to the radial composition
structure which changes over time, while the traditional chemical composition
refers to the uniform abundance at the time of formation.) To fix the
transformation correctly, it is necessary that a star of a given mass, age,
and surface abundance have the appropriate color and $M_V$. Once this scale is
fixed, assuming that the theoretical models are a close approximation to
reality, the surface properties of any star of a given mass, age, and
composition may be derived and the results used to constrain the properties
of individual stars and/or clusters.

The object used to fix the zero point of the models is
invariably the sun, i.e., a solar mass star with an age of 4.6 Gyr and the
assumed chemical composition of the sun should have solar luminosity and
temperature. The solar composition contains some flexibility in that while
$Z$, the metal fraction by mass, usually lies between 0.016 and 0.018, 
$Y$, the helium fraction, ranges from 0.27 to 0.30. Ultimately, these modest
variations have little effect if the solar-mass model is appropriately
zeroed and the remaining comparisons are done differentially. Differences
in the implied zero points of the models, as well as differences in the
input physics, have contributed to some of the scatter in derived ages
and moduli for clusters over the years. Despite the improvement in the
models and isochrone morphology through better opacities and the 
inclusion of convective overshoot, problems remain in setting the absolute
scale. 

First, there are still differences in the transformations between 
the theoretical and observational plane. Though this issue has been noted
a number of times over the last decade, it will be reexamined because
it is important to determine if the effect is uniform at all [Fe/H]. 

Second, a subtle but non-negligible
problem arises because of the production of isochrones at specific abundances.
The non-overshoot isochrones of VandenBerg (1985)\markcite{v1} were 
developed at $Y$ = 0.25
and $Z$ = 0.0169 (solar), 0.010, 0.006, 0.003, and 0.0017. The Bertelli et al.
(1994\markcite{b2}; hereinafter referred to as BE) models have 
($Y,Z$) = (0.352, 0.05), (0.28, 0.02), 
(0.25, 0.008), (0.24, 0.004), (0.23, 0.001) while the Schaerer et al. 
(1992)\markcite{s1} and
Schaller et al. (1993\markcite{s2}; hereinafter collectively referred to as 
GE\markcite{s1}) isochrones
have ($Y,Z$) = (0.30, 0.02) and (0.264, 0.008). The procedure used by
most observers is to adopt the isochrones closest in composition to solar
as [M/H] = 0.0 and then assign abundances differentially. For example,
$Z$ = 0.02 and 0.008 would be [M/H] = 0.00 and --0.40, respectively, for
the BE\markcite{b2} and GE\markcite{s2} isochrones. As mentioned earlier, 
such an approach would work
in an absolute sense for [Fe/H] = 0.00, even if ($Y,Z$) for the sun were
slightly different from the isochrone values, if the observational 
isochrones were zeroed to the sun at the adopted age. However, 
for other metallicities
even differential comparisons may fail because, while the appropriate value
of $Z$ is available, it is linked to a specific value of $Y$ which remains 
unknown. Clearly what is needed is a means of zeroing the scale for non-solar
compositions. With the advent of the {\it Hipparcos} catalog of
trigonometric parallaxes (Perryman et al. 1997\markcite{p2}), this 
becomes possible though, to
date, the primary focus has been on the subdwarfs due to their link to
globular clusters (e.g., Reid 1997\markcite{r1}; 
Gratton et al. 1997\markcite{g1}). 
In the discussion below, we will attempt to test the
isochrones for a composition more appropriate to the old disk, [Fe/H] = --0.4,
and the LMC. 

\subsection{Solar Isochrones}
The first and most straightforward step is the zeroing of the solar isochrones.
Though some disagreement still remains regarding the exact value of the
solar color, the range commonly adopted is $(B-V)$ between 0.64 and 0.66;
over the last decade, we have consistently used $(B-V)$ = 0.65 with $M_V$ 
of 4.84 (e.g., Anthony-Twarog et al. 1990\markcite{a1}; Daniel et al.
1993\markcite{d1}; 
Anthony-Twarog et al. 1994\markcite{a3}; 
Twarog et al. 1995\markcite{t3}).  Thus,
a solar mass star with solar composition at 4.6 Gyr should have $(B-V)$ and
$M_V$ of 0.65 and 4.84, respectively. Using the isochrones in the observational
plane as published for BE\markcite{b2} or constructed and interpolated with the
programs of Meynet et al. (1993)\markcite{m7} for GE\markcite{s1}, 
one can derive the 
$(B-V)$, $M_V$ for a solar mass star at 4.6 Gyr for the two compositions
which bracket the potential solar abundance, $Z$ = 0.02 and $Z$ = 0.008.
We have also included the same information for the $Z$ = 0.05 models of
BE\markcite{b2}.
The results are illustrated in Fig. 1 where the open circle represents
the true sun, squares and triangles are data for BE\markcite{b2} and 
GE\markcite{s2}, respectively.
\placefigure{fi1}

It is apparent from the trend that the BE\markcite{b2} compositions do bracket
the solar position. One could use the BE\markcite{b2} isochrones with only
minor adjustment by fixing $Z$ = 0.018 as the true solar value, a plausible
choice which is consistent with the claims that the BE\markcite{b2} 
models accurately reproduce the solar characteristics in the 
theoretical plane and that the transformations to the observational plane
are reliable. In the case
of GE\markcite{s1} no simple interpolation will match the solar properties. 
Unfortunately,
in both cases, the common procedure adopted by most investigators when
comparing the isochrones to clusters is to adopt 
the $Z$ = 0.02 isochrones
as the solar value, thereby incorporating a zero-point error in all the
solar isochrones. To rezero the $Z$ = 0.02 isochrones as solar, 
one needs to add the 
offsets [$\Delta$$V$, $\Delta$$(B-V)$] = [--0.07, +0.045] to the isochrones
of GE\markcite{s2} and [--0.04, --0.032] to those of BE\markcite{b2}. 
It should be emphasized that
offsets of this size are neither new nor unexpected. Adjustments to the
VandenBerg (1985)\markcite{v1} scale have been discussed in 
Twarog \& Anthony-Twarog 
(1989)\markcite{t1}
and Anthony-Twarog et al. (1990)\markcite{a1} and to the Castellani et al. 
(1992)\markcite{c1} scale and
the Maeder \& Meynet (1991)\markcite{m1} scale in Twarog et al. 
(1993)\markcite{t4}. That these 
adjustments are not the product of errors and/or differences in the
models nor unique to the two sets included in this investigation 
may be seen in Figs. 10 and 11 of Nordstr\"{o}m et al. (1997)\markcite{n1} and 
Figs. 5 and 10 of Mermilliod et al. (1996)\markcite{m6}.
In the theoretical plane the unevolved main sequences exhibit 
minor differences; in the observational plane on the unevolved main
sequence at $(B-V)$ = 0.7, $M_V$ differs by 0.5 mag, with the models 
of BE\markcite{b2} being brighter.

If we apply the offsets above to the $Z$ = 0.02 isochrones, one gets the 
comparison
seen in Fig. 2 for an age of 4 Gyr. The unevolved main sequences are
identical. The color at the turnoff and on the giant branch is redder
for GE\markcite{s1}, while the luminosity of the subgiant branch is brighter.
We close the discussion of the solar models by noting that the derived
offsets are
important not just because they are of significant size but because
they are systematic. Without correction, main sequence fitting with the
GE\markcite{s2} isochrones will produce distance moduli which are uniformly 
too small by 0.35 mag;
the BE\markcite{b2} isochrones will produce moduli which 
are too large by 0.15 mag.
Ages tied to the color of the turnoff will be too 
old in the GE\markcite{s1} case and
slightly too young in the BE\markcite{b2} case.
\placefigure{fi2}

\subsection{Isochrones of Intermediate Metallicity}
Given the offsets derived for the solar models, the obvious question is:
should the same offsets apply to isochrones of non-solar composition?
In the past, due to the lack of any additional means of testing the
isochrones, the default answer has been yes. The problem this presents is
illustrated in Fig. 3 where we have superposed the BE\markcite{b2} and 
GE\markcite{s1} isochrones
for an age of 2 Gyr and [Fe/H] = --0.4. In the theoretical plane, the
subgiant branches and the unevolved main sequence have identical luminosities.
At a given age, the GE\markcite{s2} isochrones extend 
toward higher log $T_e$ at the
turnoff and lower log $T_e$ on the giant branch. Using the BE\markcite{b2} 
and GE\markcite{s2} 
transformation to the
observational plane as shown in Fig. 4, one would expect a differential offset
comparable to that found for the solar isochrones, i.e., at a given $(B-V)$,
the main sequences should differ by approximately 0.5 mag. Clearly,
they do not; the typical offset in $M_V$ at a given $(B-V)$ 
on the unevolved main
sequence is only 0.12 mag, with the GE\markcite{s1} isochrones being fainter. 
Application of the same offsets as derived for solar abundance models will
lead to a differential of 0.35 mag in $M_V$ at a given $(B-V)$ 
on the main sequence,
with the isochrones of GE\markcite{s2} being brighter. 
\placefigure{fi3}

Which transformation between the theoretical and observational plane, if
any, is correct? Moreover, even if the transformations coincided, do the
isochrones truly reproduce stars in the solar neighborhood with [Fe/H]
= --0.4, i.e., is the correlated change in helium and metals appropriate?
The latter question is significantly more challenging in that it requires
knowledge of the mass of an unevolved star with independently derived 
$M_{bol}$, log $T_e$, and [Fe/H]. A simpler approach can resolve 
the former question
while empirically sidestepping the latter issue: require the unevolved
main sequence CMD of the isochrones to coincide with the cooler dwarfs with
[Fe/H] = --0.4 in the solar neighborhood as defined by the {\it Hipparcos} data.
\placefigure{fi4}

\subsection{The {\it Hipparcos} Sample}
To isolate the metallicity range of interest we make use of the $uvby$ 
catalog of G stars compiled by Olsen (1993)\markcite{o1}. For cooler dwarfs, $uvby$
photometry can supply the metallicity via the calibration derived by
Schuster \& Nissen (1989)\markcite{s3}; the data of 
Olsen (1993)\markcite{o1} have been adjusted
slightly using the relations supplied by Olsen (1993)\markcite{o1} to place them on
the same system as the calibration stars. Stars in the $(b-y)$ range from
0.39 to 0.59, $V$ $<$ 9.0, [Fe/H] between --0.3 and --0.5 and classed as dwarfs
for transformation purposes by Olsen (1993)\markcite{o1} were isolated. The {\it Hipparcos}
parallax
sample was searched and all stars with measurable parallax were identified. 
Of these, the sample was further restricted to all stars with $\sigma$/$\pi$
$\leq$ 0.1.

Though $V$ and $(B-V)$ were available from the {\it Hipparcos Catalog}, the
quoted errors in $(B-V)$ covered a non-negligible range, in contrast with
the $(b-y)$ values of Olsen (1993)\markcite{o1}. To maximize the internal consistency
of the data, a linear transformation was derived between $(b-y)$ and $(B-V)$,
weighting the data by the inverse of the error in $(B-V)$. From
137 stars over the $(b-y)$ range of interest, $(b-y)$ was transformed
to $(B-V)$ using

\centerline{$(B-V)$ = 1.873 ($\pm$ 0.032) $(b-y)$ - 0.115 ($\pm$ 0.014)}

The rms scatter about the mean relation is only $\pm$ 0.014 mag with no
dependence on [Fe/H].

Distances and $M_V$ were derived for each star from the parallaxes after
applying corrections for the Lutz-Kelker effect (Lutz \& Kelker 
1973\markcite{l1}; Koen
1992\markcite{k2}). All corrections were less than 0.1 mag. The resulting CMD is presented
in Fig. 5 where the error bars are based upon the one sigma error in parallax.
To improve clarity, the error bars in $(B-V)$ are not drawn; they are
very similar for all the stars and typically less than $\pm$ 0.01. As one might 
expect, the error bars for the parallaxes 
are smallest for the coolest and intrinsically faintest dwarfs.
Superposed is the zero-age-main-sequence (ZAMS) for [Fe/H] = --0.4 from
the models of BE\markcite{b2}.
\placefigure{fi5}

A significant number of stars are positioned well above the unevolved, cool
main sequence. These are expected to fall into two general classes. The
brightest stars ($M_V$ $<$ 4.2) are subgiants evolving across the 
HR diagram and up the
giant branch for the first time. The stars at intermediate luminosity that
parallel the main sequence may
be stars with larger than average errors in their parallaxes, but the
more likely possibility is that they are binaries, shifted above the main
sequence by the presence of a companion of comparable brightness. 
As a check on this interpretation one can use the $uvby$ data of Olsen (1993)\markcite{o1}.
If the stars that parallel the main sequence are binaries, their indices should
resemble the stars on the ZAMS since the binaries should be composed of two
similar stars. In contrast, the subgiants with lower log $g$ should exhibit
$c_1$ indices distinct from the main sequence stars. This is exactly
what is seen in Fig. 6, where the stars redder than $(B-V)$ = 0.65 have 
been sorted into two groups, those brighter and those fainter than $M_V$
= 4.2. The crosses represent subgiants while the squares are the unevolved
stars and the probable binaries. For $(B-V)$ $>$ 0.71, the separation is
virtually perfect. The one star classed as a binary which falls among the
subgiants is HD 174429, a chromospherically active member of the Pleiades
supercluster, probably a pre-main-sequence star 
(Eggen 1995\markcite{e1}; Henry et al. 1996\markcite{h1}). Its inclusion in the sample
is undoubtedly the mistaken product of anomalous indices induced by its 
active nature. Though the change in $c_1$ is modest given the change in
$M_V$, it is encouraging to see that $c_1$ does have the capability to 
distinguish between dwarfs and subgiants at cooler temperatures for 
stars at metallicities normally associated with the disk, not just
for subdwarfs as in Pilachowski et al. (1993)\markcite{p3}.
\placefigure{fi6}

Note also that the range in $c_1$ increases for supposedly unevolved stars
blueward of $(B-V)$ = 0.7. This sharp change in sensitivity also occurs
with $hk$ photometry (Anthony-Twarog et al. 1991\markcite{a2}; 
Twarog \& Anthony-Twarog
1995\markcite{t2}), as illustrated in Eggen (1997)\markcite{e2}. Since the $Ca$ and $u$ filters are
located shortward of 4000 \AA, it may indicate a significant change in
the source of ultraviolet continuum opacity near $(b-y)$ = 0.45.

To test the match between the field stars and the isochrones, only the
redder stars ($(B-V)$ $\geq$ 0.7) will be included. As
one moves up the main sequence to $(B-V)$ $<$ 0.7, the errors increase
while the possibility grows of finding stars evolving off the main sequence
near the hydrogen-exhaustion phase. The $M_V$, $(B-V)$ relation
for the unevolved main sequence at [Fe/H] = --0.4 and $(B-V)$ $\geq$ 0.7 was
taken from BE\markcite{b2} and the distance above or below the main sequence was 
measured for each star. From Fig. 5, it is clear that the single stars
with small errors scatter uniformly within $\pm$ 0.3 mag of the main sequence.
Restricting the sample to stars within 0.3 mag of the main sequence and
weighting the points by the inverse error in $M_V$, one finds an average
offset between the parallax stars and the isochrone relation of less than
--0.005 mag; a histogram of the weighted distribution in $\Delta$$M_V$
using all the points confirms that the BE\markcite{b2} isochrone relation for
[Fe/H] = --0.4 is on the same
scale as the parallax stars with an uncertainty of less than $\pm$0.02 mag.
The weighted mean of the metallicity of the stars used in defining the
zero point of the unevolved main sequence is [Fe/H] = --0.38.

In sharp contrast with the solar models, {\it no offset is required}. Moreover,
there is no evidence for a metallicity dependence in the location of the
ZAMS with the range of points from [Fe/H] = --0.3
to --0.5, indicating that the modest change expected is washed out by the
combination of errors in [Fe/H], $(B-V)$, and $M_V$. 

How much of a shift
should we expect? If we superpose the ZAMS for the isochrones at [Fe/H]
= 0.0 and --0.4, the metal-poor ZAMS is, as expected, fainter by 0.4 mag.
If the abundance effect is approximated by a linear trend, this implies
that $\Delta$$M_V$/$\Delta$[Fe/H] = --1.0. If the sample is divided in two
groups between [Fe/H] = --0.3 and --0.39 and between --0.4 and --0.49, the
expected difference in $M_V$ at a given $(B-V)$ is only 0.1 mag. Reid 
(1998)\markcite{r2} has discussed the trend with [Fe/H] for {\it Hipparcos} stars
with [Fe/H] between --0.4 and --2 and finds a slope at the metal-rich end
of the scale consistent with theory, but the data scatter is large and the stars
lie between $(B-V)$ = 0.55 and 0.75, bluer than the current sample.

Since the shift in $M_V$ for GE\markcite{s1} relative to the 
isochrones of BE\markcite{b2} is well determined (see Fig. 4)
and the slope of the main sequence is known, one can easily show that an
offset of $\Delta$$(B-V)$ = +0.022 is required to place the GE\markcite{s2} isochrones
at [Fe/H] = --0.4 on the {\it Hipparcos} scale, the same direction as for
the solar isochrones, but smaller. The solution of adjusting the scale 
by shifting only $(B-V)$ reiterates a point made at
the start of this comparison. Because we do not have mass and/or age
information for the stars being used to zero the scale, we cannot say if
a shift is required in $(B-V)$, $M_V$, or both. We have taken the simplest
approach of applying the entire offset in $(B-V)$ because the log $T_e$ to
$(B-V)$ transformation is generally considered the most uncertain. It should
be emphasized that the zero point of the age scale, fixed via the mass,
remains unknown.

Before the comparisons are made between the clusters and the isochrones,
the question of the relative metallicity scales should be addressed. The
abundances of the clusters are tied to the revised DDO calibration for giants by
Twarog \& Anthony-Twarog (1996)\markcite{t6}, but the field dwarf abundances
are defined by the Schuster \& Nissen (1989)\markcite{s3} $uvby$ calibration.
The DDO data are ultimately linked to a composite catalog (Anthony-Twarog
\& Twarog 1998)\markcite{a4} of high-dispersion,
spectroscopic abundances from the literature, transformed to a common system
which is zeroed to an adopted Hyades [Fe/H] of +0.12 and the field
star-globular cluster scale of Sneden et al. (1991)\markcite{s7} 
and Kraft et al. (1992)\markcite{k3}. 

Though hardly ideal, one method
for linking the dwarf and giant abundance scales is through the recent
revisions of both systems by Carretta \& Gratton (1997)\markcite{c4}
and Clementini et al. (1998)\markcite{c5}. On the revised system, 
the difference in [Fe/H] for giants in the globular
clusters common to both, in the sense (CG - SKPL), is +0.11 $\pm$ 0.02 (s.e.m.)
for 7 clusters ranging from [Fe/H] = --0.7 to --2.2. For the dwarfs, Clementini
et al. (1998)\markcite{c5} find a difference in [Fe/H] among the stars 
common to Schuster
\& Nissen (1989)\markcite{s3}, in the sense (CGCS - SN), of +0.10 $\pm$ 0.01
(s.e.m.). If the revised spectroscopic systems are identical, this implies
that the data of SN and the revised DDO scale are virtually identical.

A more direct approach would use a comparison between the spectroscopic catalog
which defines the DDO calibration and the stars common to the {\it uvby} sample
of Fig. 5 or the spectroscopic dwarf catalog of Clementini et al.
(1998)\markcite{c5}. Though the spectroscopic catalog compiled by
Anthony-Twarog \& Twarog (1998)\markcite{a4} is dominated by giants, some
surveys included a number of subgiants, and an even smaller fraction included
modest numbers of dwarfs. Because of this, the overlap with the current
dwarf sample is small, 3 stars for the {\it uvby} data and 7 for the
Clementini et al. (1998)\markcite{c5} data. The scatter is significant,
but the conclusions from both comparisons are that the {\it uvby} scale is
too metal-rich compared to the DDO scale by between 0.1 and 0.2 dex. Taken
at face value, this implies that the stars used to zero the intermediate
isochrones are, on average, closer to [Fe/H] = --0.55 than --0.4 and, thus,
the unevolved main sequence is too faint by approximately 0.15 mag. However,
given the small sample and the questionable application to dwarfs 
of the transformation equations derived for giants, we will assume that
the DDO and {\it uvby} scales have the same zero point.  

\section{The Clusters: NGC 2420 and NGC 2506}
\subsection{Basic Cluster Parameters}
Though there are over a dozen clusters in the galactic anticenter with
abundances ranging from [Fe/H] = --0.2 to --0.6 
(Twarog et al. 1997\markcite{t5}), 
we will restrict our
isochrone comparison to only two, NGC 2420 and NGC 2506. These two clusters,
as demonstrated below, have virtually identical CMDs but, more important,
the level of information available for stars within these clusters is
unique among the anticenter sample. There is excellent CCD and photographic
broad-band photometry, intermediate-band photometry for reddening and
abundances, spectroscopy for abundance estimation and radial-velocity
determination, and proper-motion analysis for membership determination.
The membership estimation is especially critical given the sparsely populated
red giant branches, in contrast with the clusters regularly found in the
Magellanic Clouds. Inclusion of only a modest number of field interlopers
can skew the derivation of the color and luminosity of the red giant branch
and clump.

A detailed discussion of the photometry and CMD for NGC 2420 can be found in 
Anthony-Twarog et al. (1990)\markcite{a1}. For our sample we will 
include all stars
classified as definite and probable members, using both CCD and photographic
data as presented in Anthony-Twarog et al. (1990)\markcite{a1}. The 
metallicity of the
cluster, as derived in Twarog et al. (1997)\markcite{t5}, is 
[Fe/H] = --0.28, assuming $E(B-V)$ = 0.05.

For NGC 2506, primary use is made of the $UBVRI$ CCD survey by Marconi et al.
(1997)\markcite{m2}. Because this sample does not cover the entire cluster, it has
been supplemented by the photographic data of McClure et al. (1981)\markcite{m5}. To
ensure that they are on the same system, a comparison has been made between
the CCD data and the photographic for 106 stars with $V$ $<$ 15.10 common
to the two surveys. This range has been chosen because it includes the 
giant branch, subgiant branch, and the top of the main sequence turnoff.
It is found that over this magnitude range, the $V$ magnitudes are on
the same system within $\pm$0.01 mag, if the extreme deviants are
removed, as confirmed by the plot in Fig. 4 of 
Marconi et al. (1997)\markcite{m2}; the
scatter about the mean is typically $\pm$0.03 mag. The deviant stars are
relevant because their distribution in $\Delta$$V$ is asymmetric; the
majority are too bright in the photographic survey. Though some of the 
stars with large residuals might be variables, a more mundane explanation
is provided by the cluster star chart: the sample of deviants is dominated
by stars with probable contamination by stars of the same or brighter
magnitude. With photographic aperture photometry one can minimize contamination
by offsetting the star in the aperture, but, in extreme cases, it is
impossible to remove. This solution is confirmed by the comparison in $(B-V)$,
where the scatter in the residuals at a given color is the same as that
found for $V$, while many of the deviants in $V$ show no anomaly in $(B-V)$.
The one difference with $(B-V)$ is that a small color term does appear 
among the residuals between the CCD
and photographic data in $(B-V)$. The photographic data were transformed to
the CCD system using the relation
\medskip
\centerline{$(B-V)_{CCD}$ = 0.97 $(B-V)_{PG}$ + 0.01}
\medskip
All stars from McClure et al. (1981)\markcite{m5} with membership 
probability greater than
or equal to 80\% (Chiu et al. 1981)\markcite{c2} and $V$ brighter than 15.1 not 
included in the CCD work of
Marconi et al. (1997)\markcite{m2} were identified and their 
photometry transferred to
the CCD system. Their locations within the cluster were
checked and, if they suffered from potential contamination by a nearby star
of comparable brightness, they were tagged. The photograpic data were merged
with the CCD data for cluster members. 

Based upon the same combination of DDO photometry and spectroscopy as
used for NGC 2420, Twarog et al.
(1997)\markcite{t5} found the metal
abundance of NGC 2506 to be --0.38, adopting $E(B-V)$ = 0.05.
Because
of the change in $(B-V)$ introduced by the transformation between the 
photographic data and the CCD data, we have rederived the reddening estimate
for NGC 2506 using the DDO approach of Janes (1977)\markcite{j1}. If all 10 stars within
the calibration limits are included, $E(B-V)$ = 0.02 $\pm$ 0.05 (s.d.). If
stars with membership probabilities below 70\% are removed, the mean
reddening for 6 stars drops to 0.00 $\pm$ 0.02 (s.d.). These results are
consistent with the original estimates of $E(B-V)$ = 0.05 and 0.03, 
respectively,
by McClure et al. (1981)\markcite{m5} 
in that the
color transformation derived above produces a shift of between 0.02 and 0.03
mag in $(B-V)$ for the red giants with the CCD system being bluer. Thus,
with the same DDO indices, the previous reddening estimates are 
reduced by the same amount. Note
that this implies that the absolute color of the giant branch is approximately
the same irrespective of which photometric system is adopted.

As a check for consistency, we reanalyzed the DDO photometry for NGC 2420
(McClure et al. 1974)\markcite{m3} using the 
Janes (1977)\markcite{j1} approach; the earlier discussion
used a technique outlined by McClure \& Racine (1969)\markcite{m4}. 
From 11 stars, $E(B-V)$
= 0.02 $\pm$ 0.03 (s.d.); the result remains unchanged if the one non-member
is dropped. This result is consistent with the extensive discussion of the
relative reddening of these clusters found in Anthony-Twarog et al.
(1990)\markcite{a1}.
The two clusters have very similar reddenings, but the absolute reddening 
of the two clusters
remains more uncertain. Indications from the field stars 
observed with the clusters and
the reddening maps of Burstein \& Heiles (1982)\markcite{b3} are 
that the true reddening is
slightly higher than that derived from the DDO sample, consistent with the
giants being bluer than expected due to their lower [Fe/H] relative to the
field star sample, though Janes (1977)\markcite{j1} claims that metallicity effects on
the DDO technique are small. The one significant addition to the question
since Anthony-Twarog et al. (1990) is that of Schlegel et al. 
(1998\markcite{s6}).
The reddening maps of Schlegel et al. (1998\markcite{s6}) produce 
$E(B-V)$ = 0.04
for NGC 2420, consistent with the results of 
Burstein \& Heiles (1982)\markcite{b3},
and 0.08 for NGC 2506, smaller than the 
Burstein \& Heiles (1982)\markcite{b3} value of 0.12, but
indicative of a higher $E(B-V)$ for NGC 2506.
For purposes of discussion, we will assume, as in Twarog et al. 
(1997)\markcite{t5}, that the two clusters have the same, but slightly lower,
reddening,
$E(B-V)$ = 0.04. In addition to the consistency of the reddening
maps, we give more weight to the DDO result for NGC 2420 because of an
apparent problem with the CCD photometry of Marconi et al. (1997)\markcite{m2},
an issue we will return to in Sec. 3.3.
Given this minor 
change in reddening, the difference in [Fe/H]
between NGC 2420 and NGC 2506 remains 0.10 dex, with NGC 2420 being more
metal-rich; on the system of Twarog et al. (1997)\markcite{t5}, 
NGC 2506 has [Fe/H] = --0.39.

As a final step before deriving the absolute moduli of the two clusters, we
make an additional comparison to constrain the cluster parameters, 
a differential
comparison of the two clusters, independent of the isochrones. This is 
especially
useful because the unevolved main sequence of NGC 2420 is much better defined 
than that of NGC 2506.
Since the differential metallicity is known, one can estimate that the
unevolved main sequences of the two clusters should differ by approximately 0.1
mag, with NGC 2420 being brighter. Additionally, the red giant 
branch of NGC 2420
should be redder by 0.01 to 0.02 mag, assuming they have the same age 
(see Sec. 3.2).
As a secondary constraint, we assume that they differ in reddening by 0.00 $\pm$
0.02 mag. Within these constraints, the optimal displacements, 
in the sense (2506-2420),
are --0.01 in $(B-V)$ and +0.50 in $V$, implying that the reddening for
NGC 2506 should be {\it smaller} than that for NGC 2420. If we require 
$\Delta$$(B-V)$ = --0.02,
the main sequence condition is met if $\Delta$$V$ = 0.45, but the red giant
branches are virtually aligned. Increasing the reddening differential forces
smaller shifts in $V$, but increasingly fails to match the red giant branch.
If we lower the differential reddening to 0.0, $\Delta$$V$ = 0.55 and the giant
branches differ in color by approximately 0.03 mag, with NGC 2420 being
redder. 

An uncertainty which arises in this analysis is the exact zero point of
the CCD photometry in $(B-V)$. As pointed out above, a comparison between
the CCD data and the photographic data tied to internal cluster standards
leads to a small color term in the residuals in $(B-V)$, in that the CCD
data are too blue. If the photographic zero point is correct, the giant
branch is redder but the reddening value is higher and consistent with
the assumed relative reddening for the two clusters. Thus, a differential
offset of --0.01 in $(B-V)$ using the CCD system is equivalent to an
offset of +0.01 for the photographic system. We conclude that the ideal
differential fit for the two clusters, tied to the CCD data, is $\Delta$$(B-V)$
= --0.01 $\pm$ 0.01 and $\Delta$$V$ = 0.50 $\pm$ 0.05.

\subsection{Isochrone Fits: Distances and Ages}
Given $E(B-V)$ = 0.04 and adjusting the isochrones by --0.1 mag in $M_V$ to
account for the higher metallicity, the optimum fit of NGC 2420 to the 
intermediate metallicity isochrones of BE\markcite{b2} is shown in Fig.
7. The apparent
modulus is $(m-M)$ = 12.15. The shape of the isochrones from the main 
sequence near $(B-V)_0$ = 0.8 through the
turnoff, subgiant branch, giant branch, and clump is an excellent match to
the data, implying an age of 2.0 Gyr. Taking into account the slightly higher
metallicity of the cluster relative to the isochrones, the true age is more
like 1.9 Gyr, with an uncertainty of less than 10 \%. In fixing the distance
modulus, use has been made only of the main sequence, with special emphasis
on the color range between $(B-V)$ = 0.4 and 0.6 where the confusion caused
by field star contamination and photometric errors is minimized. Use of the
giant branch can create problems because the isochrones colors are tied to a
lower metallicity; if corrections were made to account for the higher
[Fe/H] of NGC 2420, the isochrone giant branches would be slightly redder.
\placefigure{fi7}

Fig. 8 illustrates the match for NGC 2506, adopting $E(B-V)$ = 0.04 and
$(m-M)$ = 12.70. No adjustment has been made to the isochrones because the
cluster abundance is effectively the same as the models. The match between
theory and observation is quite good near the turnoff, but the expanded 
scatter in
the main sequence below $(B-V)_0$ = 0.6 makes this region relatively useless
for fitting purposes. As expected from the differential fit discussed above,
the primary fitting region between $(B-V)_0$ = 0.4 and 0.6 is well matched
by the isochrones. The cluster is almost a perfect match to a turnoff age
of 1.8 Gyr, making it slightly younger than NGC 2420. The luminosities of
the stars on the subgiant branch are well matched by the isochrones, but it
is obvious that the giant branch of the cluster is too blue. This can be
corrected by lowering the reddening, in contradiction with the estimates
from the reddening maps, or by assuming that there is an error in color
of about 0.03 mag for the giants (see Sec. 3.3). The fit in Fig. 8 is
based upon the assumption that the latter solution is more probable.
\placefigure{fi8}

Figs. 9 and 10 show the analogous matches for the isochrones of GE\markcite{s1}.
The main sequence fits are of comparable quality to those of
BE\markcite{b2}, an expected result given the similarity between the two sets and the
fact that we have shifted the GE\markcite{s2} isochrones to match the same 
{\it Hipparcos} field stars. The GE\markcite{s1} isochrones are less optimal than those
of BE\markcite{b2} in two ways. The hydrogen-exhaustion hook at the turnoff turns too
sharply to the red relative to the clusters and the isochrone giant branches
are too red, as discussed earlier. The ages for the two clusters are 2.2 Gyr
and 2.1 Gyr for NGC 2420 and NGC 2506, respectively,
slightly older than derived from BE\markcite{b2}.  It is encouraging to 
note that 
Marconi et al. (1997)\markcite{m2} find $(m-M)_0$ = 12.5 and 
$E(B-V)$ = 0.05 for $Z$ =
0.008 for the models of BE\markcite{b2} and GE\markcite{s1}, with no 
adjustment applied to either
scale but using their own transformations between the theoretical and
observational plane. Their technique makes use of an optimum match of a 
a variety of CMD properties, including morphology and the luminosity function.
Because of the higher reddening, their ages are younger than ours. If we
were to adopt $E(B-V)$ = 0.05, then $(m-M)$ = 12.75, the same within the
errors as their 
value of 12.7.
\placefigure{fi9}

\placefigure{fi10}

\subsection{The Absolute Clump Magnitude: $V$ and $I$}
With the reddening and distance moduli in place, one can derive the intrinsic
luminosity of the clump. We have combined the giant branches for the two
clusters, including in the counts any giant within $\pm$ 0.2 mag of the
color of the red giant branch. Though it is relatively easy to 
isolate first-ascent
red giants from the clump stars because of the quality of the CMD photometry,
this approach is taken to make the distribution determination analogous to
that commonly used in less well-defined CMDs. Moreover, the distribution
function with absolute magnitude will be derived for $V$ and $I$. Though the
former is more commonly available, the latter has become the filter of
choice of late because the metallicity sensitivity of the $I$ magnitude of
the clump is believed to be substantially weaker than $V$. 

The problem with 
analyzing $I$ is that only a subset of the stars in NGC 2506 have $(V-I)$
photometry from Marconi et al. (1997)\markcite{m2}; no $I$ data are available for 
NGC 2420. We initially resolved both problems by deriving a relation between
$(B-V)$ and $(V-I)$ from the CCD photometry in NGC 2506 and applying it
to the adjusted photographic data in NGC 2506 (see Sec. 3.1) and to the
CCD data of NGC 2420. From 35 stars redder than $(B-V)$ = 0.6 and brighter
than $V$= 15.1, one finds

\centerline{$(V-I)$ = 0.91($\pm$0.03)$(B-V)$ + 0.24($\pm$0.03)} 

The rms scatter about the mean relation is only $\pm$ 0.026 mag.
The problem with this solution is illustrated in Fig. 11, where we have
plotted the $(B-V)$, $(V-I)$ data for a number of open clusters which
bracket NGC 2506 in age and metallicity; the dashed relation is the
equation listed above. The clusters included are NGC 2204 (filled
triangles), Be 39 (open circles), Mel 66 (open squares), and M67 (stars).
Data for the first three clusters are from Kassis et al. (1997)\markcite{k1};
the M67 data are from Montgomery et al. (1993)\markcite{m8}. 
Also superposed is a short dashed line indicating the shift
caused by a reddening of $E(B-V)$ = 0.10. The clusters have been adjusted
for reddening, but it is clear that unless the reddening is grossly in
error, stars will shift approximately parallel to the mean relation for
giants; no separation by metallicity is apparent for stars in the 
[Fe/H] range from 0.0 to --0.7. The relation for NGC 2506 is offset from
that of the other clusters by +0.1 in $(V-I)$ at a given $(B-V)$; the
solid line shows the NGC 2506 relation shifted by --0.1 mag in $(V-I)$. 
Whether
this offset is caused purely by a shift in $(V-I)$ or a combination of
shifts in both $(V-I)$ and $(B-V)$ cannot be decided, but a blue offset 
in $(B-V)$ at
the 0.02 to 0.03 mag level for the
CCD data relative to the photographic data for the giants is consistent
in direction with the direction of the offset seen in Fig. 11.
\placefigure{fi11}

What is the source of the discrepancy for NGC 2506 in Fig. 11? The natural
response is to conclude that a zero-point error exists in the CCD data
of Marconi et al. (1997)\markcite{m2}, possibly amounting to --0.03 mag
in $(B-V)$ and between +0.07 and +0.10 mag in $(V-I)$. Without additional 
observations,
we have no independent means of testing this conjecture and caution against
assuming that the CCD photometry, which has been tied to direct observations
of a number of standard fields, must be at fault. In particular, recent
work by Stutz et al. (1998)\markcite{s8} and Paczy\'{n}ski (1998)\markcite{p5} 
has shown that RR Lyrae
stars and red clump giants in Baade's Window appear to have anomalous
$(V-I)$ colors which cannot be explained by simple adjustments in surface
gravity or [Fe/H]. However, while one might expect discrepancies between
stars formed in galactocentric regions separated by 7 to 10 kpc, the
comparison in Fig. 11 includes clusters of comparable age and location
to NGC 2506. Thus, we have, as a matter of convenience, opted for the
assumption that an error exists in the CCD photometry for NGC 2506. We
will, however, discuss the impact if the anomalous colors in $(V-I)$ are
real.   

Combining the cluster giant branches and binning the stars as a function
of $M_V$ and $M_I$, one gets the histograms of Fig. 12 where the solid
curve is for $V$ and the dashed curve is for $I$. The peaks in the distributions
caused by the existence of the RGC are readily identified between $M_V$ =
0.25 and 0.55 and between $M_I$ = --0.65 and --0.3. One could attempt a profile
fit to the data but, given the small number of stars, we have taken a
simple average of the 22 stars between $M_I$ = --0.65 and --0.2 and $M_V$ =
0.2 and 0.6. The results are summarized in Table 1, along with the
mean color of the stars used to define the clump. Also listed are
the means under two more restrictive conditions: only stars in NGC 2506 and
only stars in NGC 2506 with CCD data.

\placefigure{fi12}

\placetable{tab1}

The RGC of NGC 2420 has the same $M_V$, within the errors, as NGC 2506,
but is redder by about 0.04 in $(B-V)$. This color differential translates
into a brighter $M_I$ and a redder $(V-I)$ for NGC 2420. If the $(B-V)$
of the giants in NGC 2506 is made redder by between 0.02 and 0.03, this
reduces the differentials between the clusters in all indicators,
consistent again with the belief that the two clusters are almost
identical in age and composition, with NGC 2420 being slightly more
metal-rich and older. If the $(B-V)$ - $(V-I)$ transformation defined
by NGC 2506 is adopted for both clusters, the average $M_I$ for the clump
is made 0.1 mag brighter and $(V-I)$ is 0.09 mag redder.
We will adopt the intermediate values of
$M_V$ = 0.47 $\pm$ 0.04 and $M_I$ = --0.48 $\pm$ 0.05, where the errors
take into account the range in options from Table 1 and the standard errors
of the mean for the individual values. Values of $M_V$ in the
range of 0.5 to 0.7 have been
obtained by Twarog et al. (1997)\markcite{t5} and 
Eggen (1998)\markcite{e3} using larger samples of
clusters with greater uncertainty in their intrinsic parameters, but both
studies find only a weak dependence of $M_V$ on [Fe/H] between --0.6 and +0.2.

For the total error budget in deriving $M_V$ and $M_I$, one must 
include the uncertainty in the fit
of the isochrones to the field stars ($\pm$ 0.02), in the cluster mean
metallicities ($\pm$ 0.02), in the definition of the
metallicity scale between the DDO and {\it uvby} systems (--0.10, +0.05),
in the fitting the cluster main sequence
($\pm$ 0.05), the uncertainty of $\pm$0.01
in the solar color ($\pm$0.055), and the $\pm$0.02 scatter 
in the allowed $E(B-V)$ for the pair 
of clusters ($\pm$0.11). It should be remembered that the differential shift
in $V$ between the clusters is correlated with the changes in $E(B-V)$ in
that a decrease in the differential reddening of the two clusters requires
a larger $\Delta$$V$.  Combining the above, the estimated errors
in $M_V$ and $M_I$ are (--0.17,+0.14) for the full sample; use of the
NGC 2506 data alone produces no significant change in the errors. Note
that the error estimate does not include any component due to the potential
zero-point error in the CCD data for $(B-V)$ or $(V-I)$. The only changes
allowed will shift the absolute magnitudes to brighter values.

\section{Applications and Future Work}
A key motivation for this investigation has been the ongoing debate on the
distances to the LMC and the SMC derived using the RGC. Though
the $M_V$ estimate is less affected by zero-point uncertainties than $M_I$, 
we will discuss
the distance based upon $I$ photometry first. A significant portion of the
debate on the giant branches has centered on the analysis
of field star samples within the two systems (e.g., Cole 1998\markcite{c3}; 
Udalski et al. 1998\markcite{u2}; Beaulieu \& Sackett 1998\markcite{b1}) and
the appropriate mixture of populations of different ages and metallicities.
As emphasized by Udalski (1998)\markcite{u1}, one can minimize the problems by
restricting the sample to
star clusters which contain stars of uniform age and abundance.
Udalski (1998)\markcite{u1} has analyzed $VI$ photometry of a 
sample of 15 star clusters in
the LMC and SMC, ranging in age from 2 to 12 Gyr. He finds rather conclusively
that there is virtually no dependence of $M_I$ on age from 2 to 10 Gyr, a 
range which
includes NGC 2420 and NGC 2506. The typical cluster sampled by
Udalski (1998)\markcite{u1} is, however, more metal poor 
than the galactic sample, 
ranging from [Fe/H] = --0.6 to --1.0 for the LMC and from --0.7 to --1.5 for
the SMC. Both Cole (1998)\markcite{c3} and Udalski (1998)\markcite{u1} 
discuss the slope of the
correction for metallicity, $\Delta$$M_I$/$\Delta$[Fe/H]; the former 
finds a value of
0.21 while the latter chooses 0.09. Adopting the middle ground, $\delta$$M_I$
= --0.06 for the LMC adjustment ([Fe/H] = --0.8) and --0.12 for the SMC
([Fe/H] = --1.2), leading to $M_I$ = --0.54 and --0.60 for the LMC and the
SMC, respectively. For the LMC clusters Udalski (1998)\markcite{u1} 
derives a mean $I_0$
of 17.88 $\pm$ 0.05 and 18.31 $\pm$ 0.07 for the SMC clusters. The resulting
distance moduli are $(m-M)_0$ = 18.42 (+0.17,--0.15) and 18.91 (+0.18,--0.16),
respectively.

These moduli are in very good agreement with the work of
Cole (1998)\markcite{c3} and marginally consistent with Udalski (1998)
\markcite{u1}; the results of
Udalski et al. (1998)\markcite{u2} can be excluded. If we adopt the anomalous
$(V-I)$ colors of NGC 2506 as being correct for both NGC 2506 and NGC 2420,
the distance moduli increase by 0.1 mag, but the question then arises as
to whether or not the MC clusters are comparable to NGC 2506.
Adopting the anomalous $(V-I)$ values as correct, the average $(V-I)_0$ color
of the RGC in NGC 2506 is 1.04, while the typical LMC cluster has $(V-I)_0$
closer to 0.9. Though NGC 2506 is more metal rich than the LMC, the anomalous
color is closer to the expected value for a giant branch of solar metallicity
as in M67 (Montgomery et al. 1993)\markcite{m8}. 

Given the moduli derived above 
one can check the distance to the LMC and the SMC using the $M_V$ of the RGC.
For six clusters in the LMC, the mean $M_V$ of the RGC is 18.78 $\pm$0.06, 
only
0.1 mag larger than the field clump identified by Beaulieu \& Sackett
(1998)\markcite{b1}.
If we choose the two youngest clusters with ages similar to NGC 2420 and
NGC 2506, the RGC mean is 18.8. For the SMC the RGC average is 19.14
$\pm$ 0.08. Assuming the moduli derived above are correct, $M_V$ = +0.36
and +0.23 for the LMC and SMC, respectively, requiring a strong
dependence of $M_V$ on metallicity. The corresponding numbers
using the Udalski (1998)\markcite{u1} moduli are +0.61 and +0.49, respectively.
Since the derived $M_V$ for NGC 2506 and NGC 2420 is +0.47 and both
clusters are significantly more metal-rich than the LMC and SMC, this
would imply no dependence of $M_V$ on metallicity.

Surprisingly, despite the many pieces that make up this chain of reasoning,
the weakest link remains the zero-point of the cluster photometry.
We emphasize, however, that in making our choices, our bias has 
been in the direction of minimizing the distance moduli by adopting the
bluer colors for $(V-I)$. 
If the zero point of the
NGC 2506 CCD photometry is confirmed and/or the giants of NGC 2420 are observed
in $VI$ and found to be consistent with NGC 2506, it would cast serious
doubt on the claim that stars of similar apparent [Fe/H] and age 
must have similar colors in $BVI$, i.e., an additional parameter must be
affecting the location of the giant branch. Moreover, it would make the distance
moduli based upon $M_I$ as derived above larger.

\acknowledgements                    
The authors are indebted to the Simbad and Vizier data access services for
the extensive bibliographic, photometric, and parallax information crucial
to the success of this investigation. Drs. Ken Janes and Monica Tosi 
graciously supplied us with files of their CCD data. The clarity of
the paper has been significantly improved due to the insightful comments
of the referee which forced us to reexamine a key assumption in an
earlier draft of the text. A.R.B. acknowledges
support of a Clyde W. Tombaugh Summer Fellowship from the Department of Physics
and Astronomy at the University of Kansas.

\newpage
\figcaption[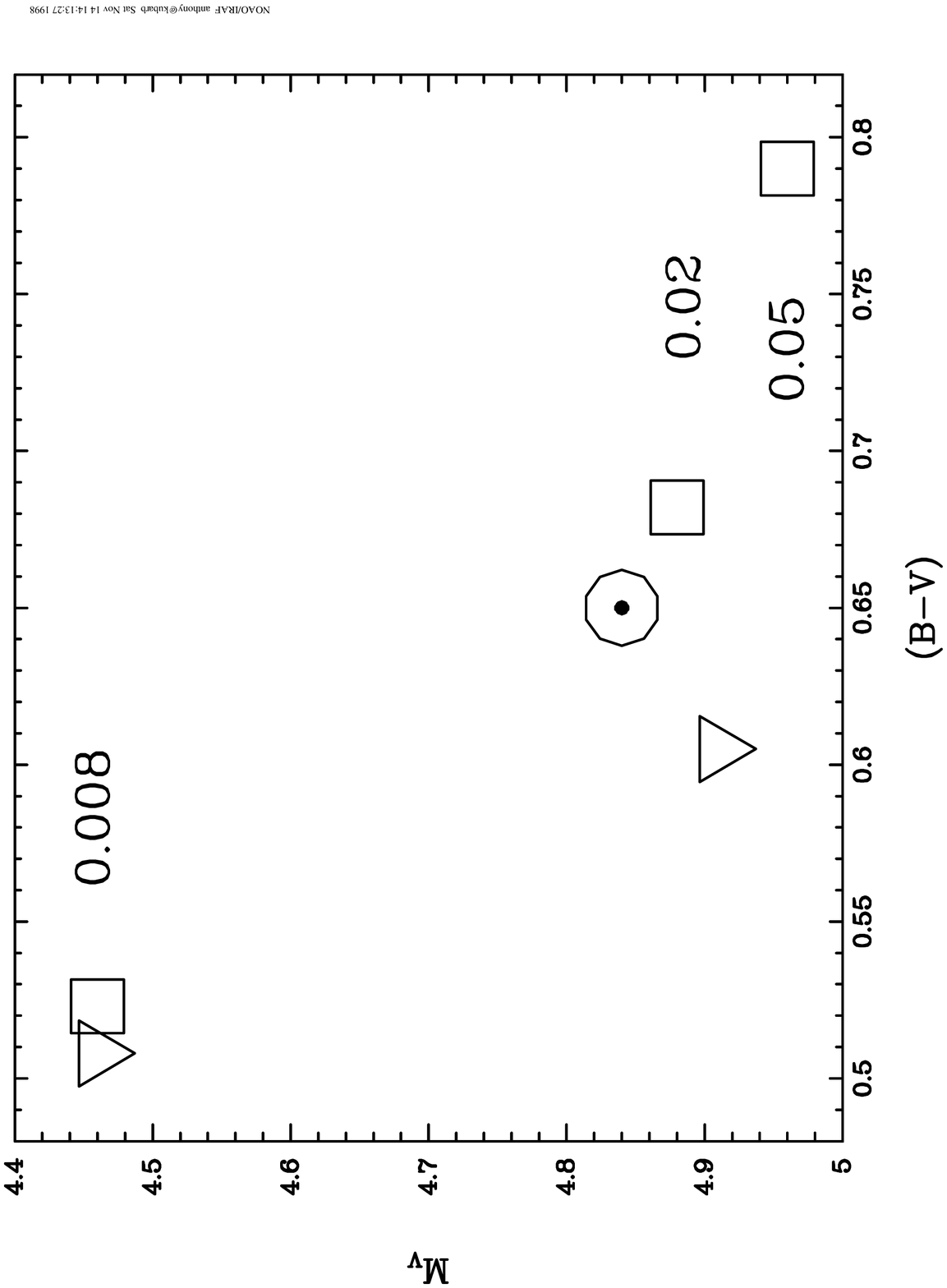]{CMD position of a star of solar mass at 4.6 Gyr for the models of
BE (squares) and GE (triangles) for different values of $Z$. The open circle is
the adopted position of the true sun.\label{fi1}}
\figcaption[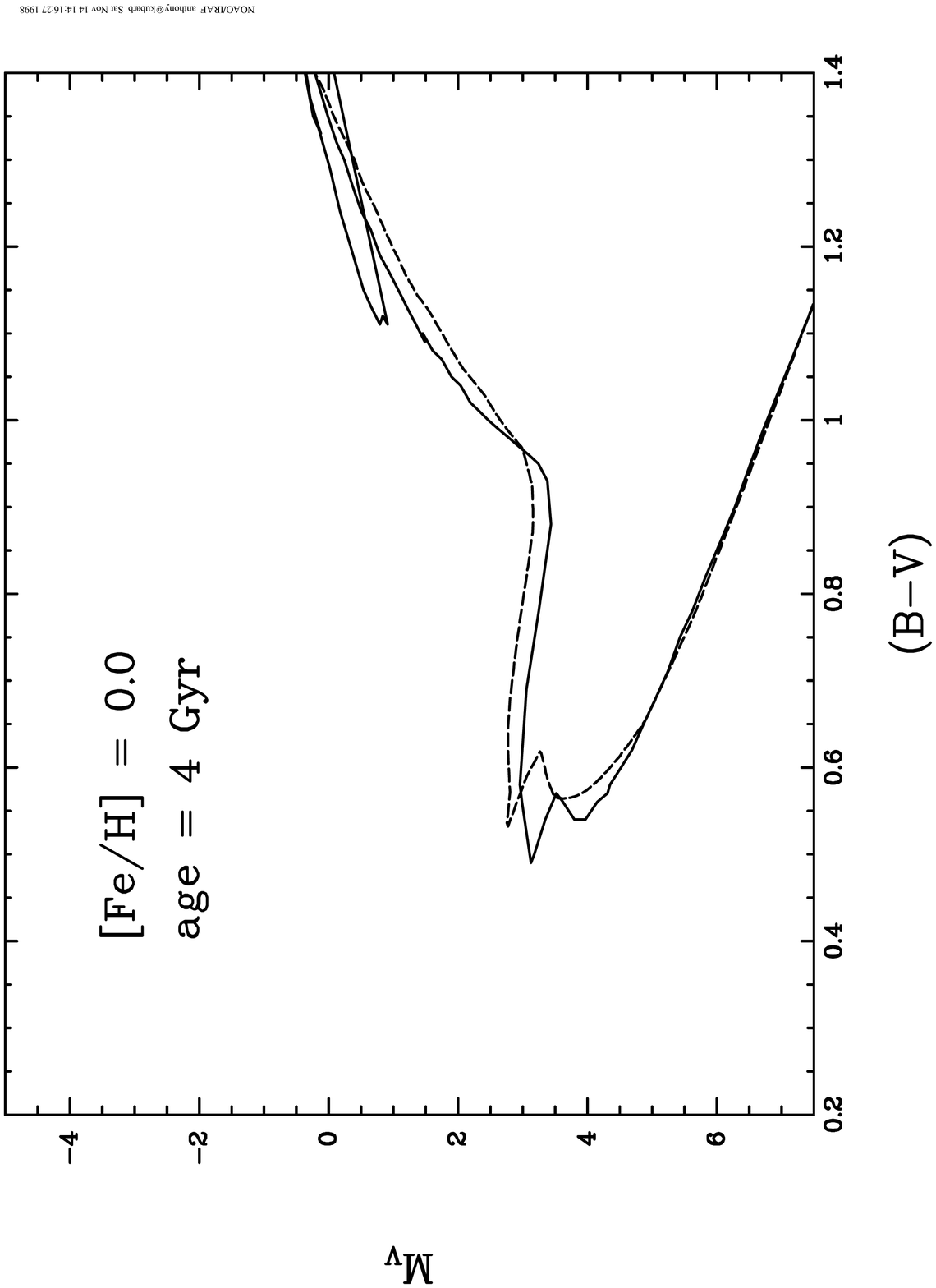]{Comparison between the 4 Gyr isochrones of BE (solid curve) 
and GE (dashed curve) at solar
abundance after offsets have been applied to rezero the scales. \label{fi2}}
\figcaption[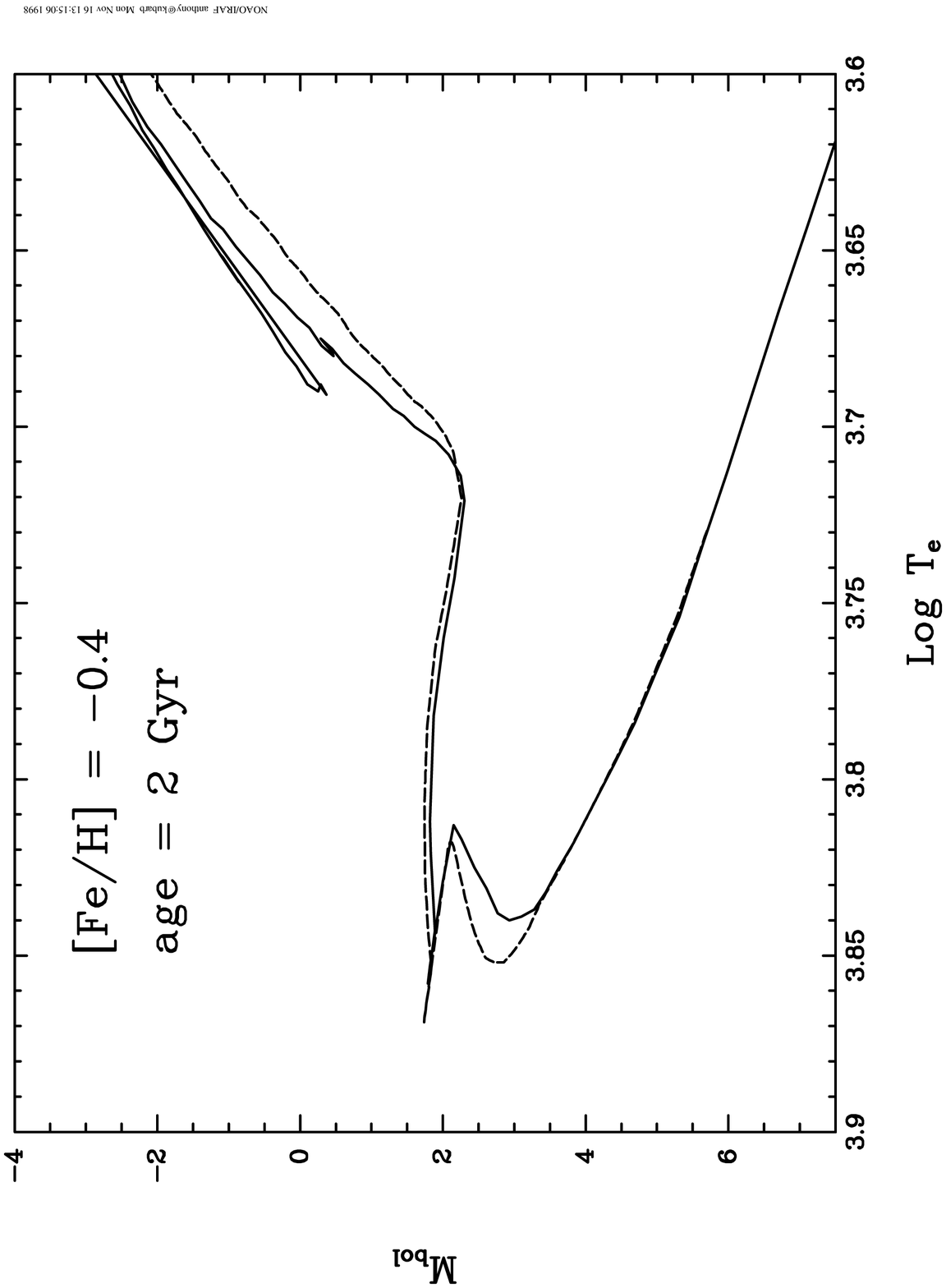]{Comparison of the 2 Gyr isochrones of BE (solid curve) and GE
(dashed curve) at intermediate metallicity in the theoretical plane. No
offsets have been applied. \label{fi3}}
\figcaption[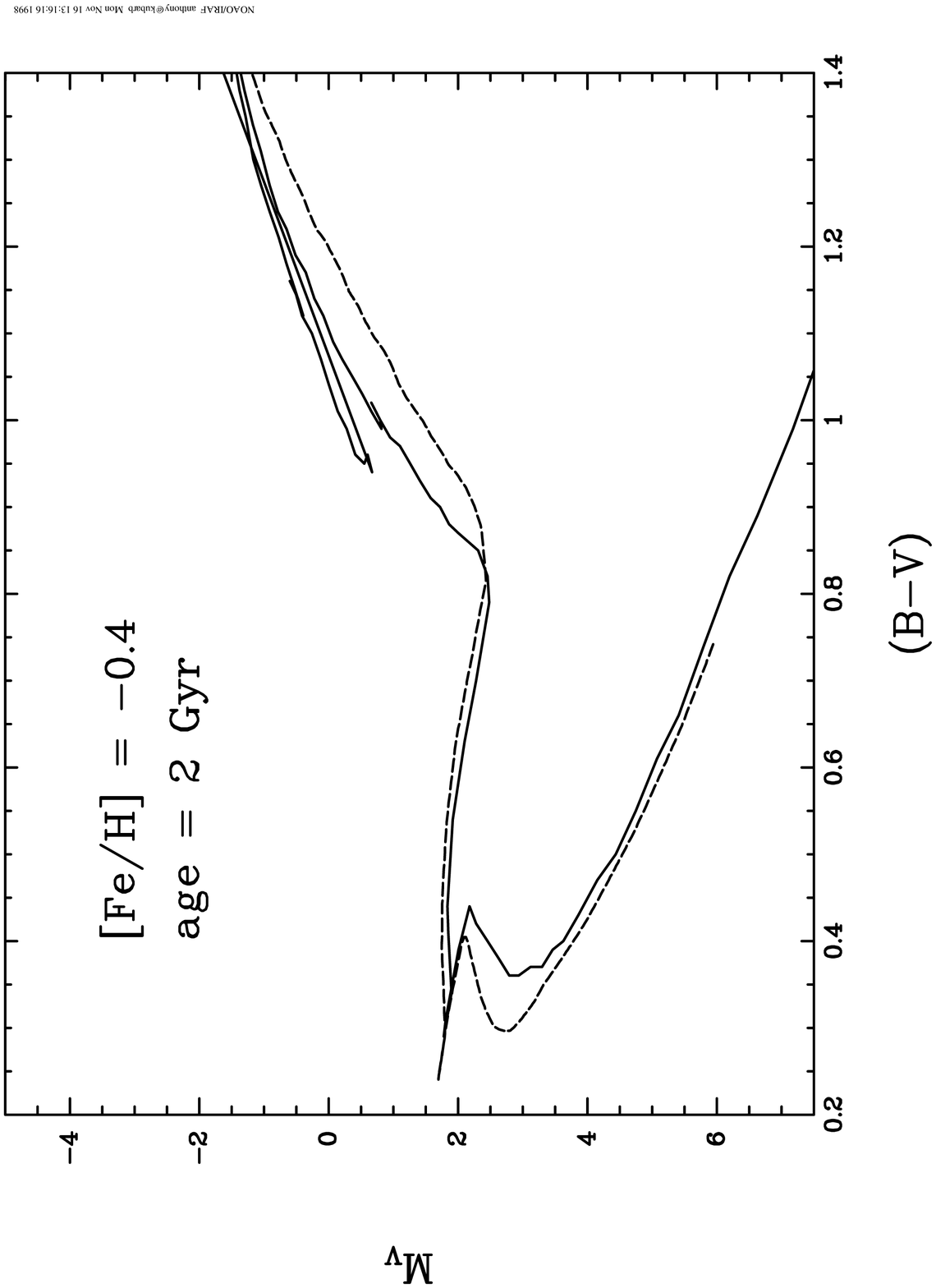]{Same as Fig. 3, but in the observational plane. \label{fi4}}
\figcaption[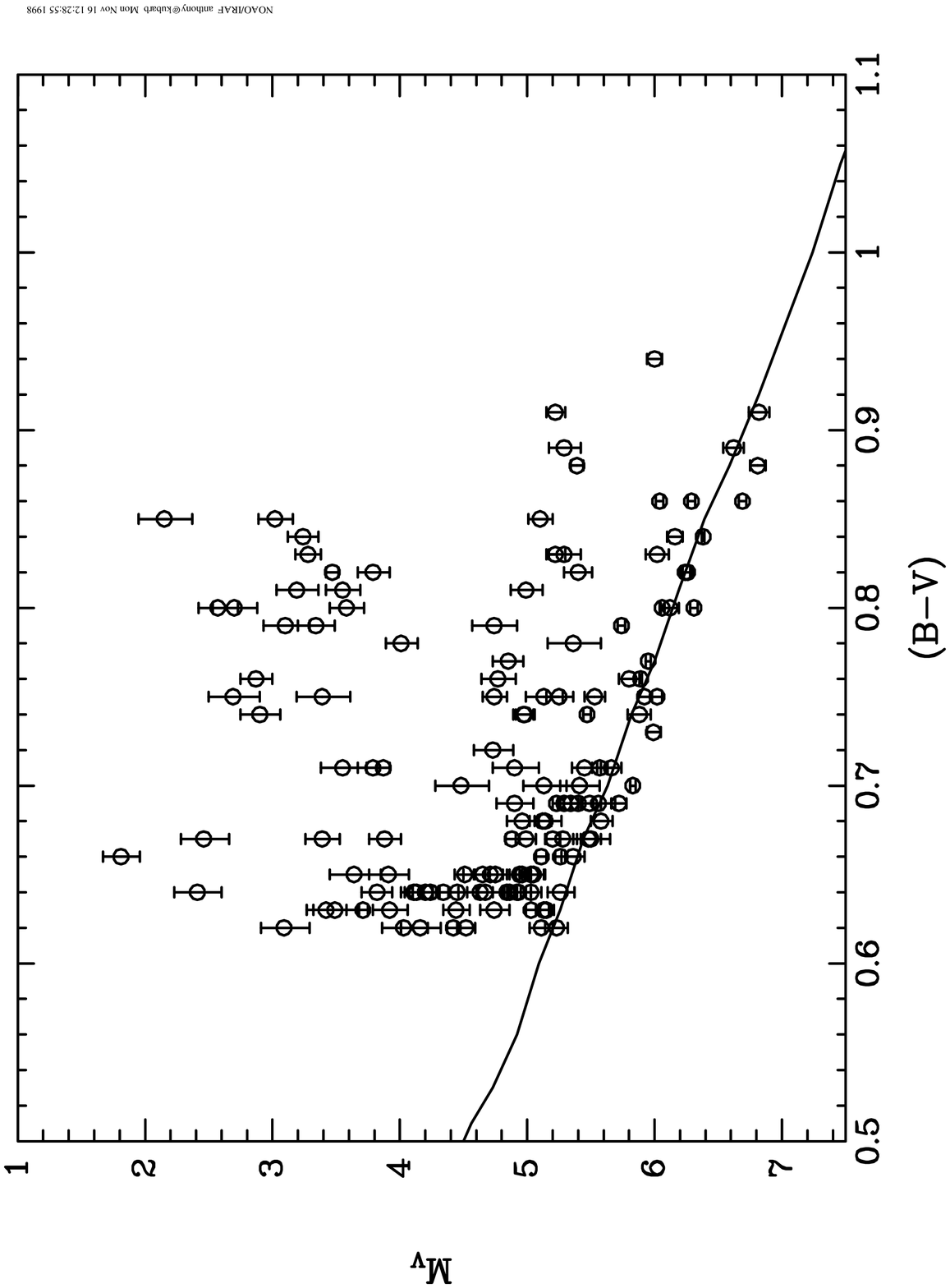]{CMD of field stars with --0.3 $\geq$ [Fe/H] $\>$ --0.5 from the
{\it Hipparcos} catalog. Error bars are one sigma errors in the parallax and
the solid line is the unevolved main sequence from the isochrones of BE for
[Fe/H] = --0.4.\label{fi5}}
\figcaption[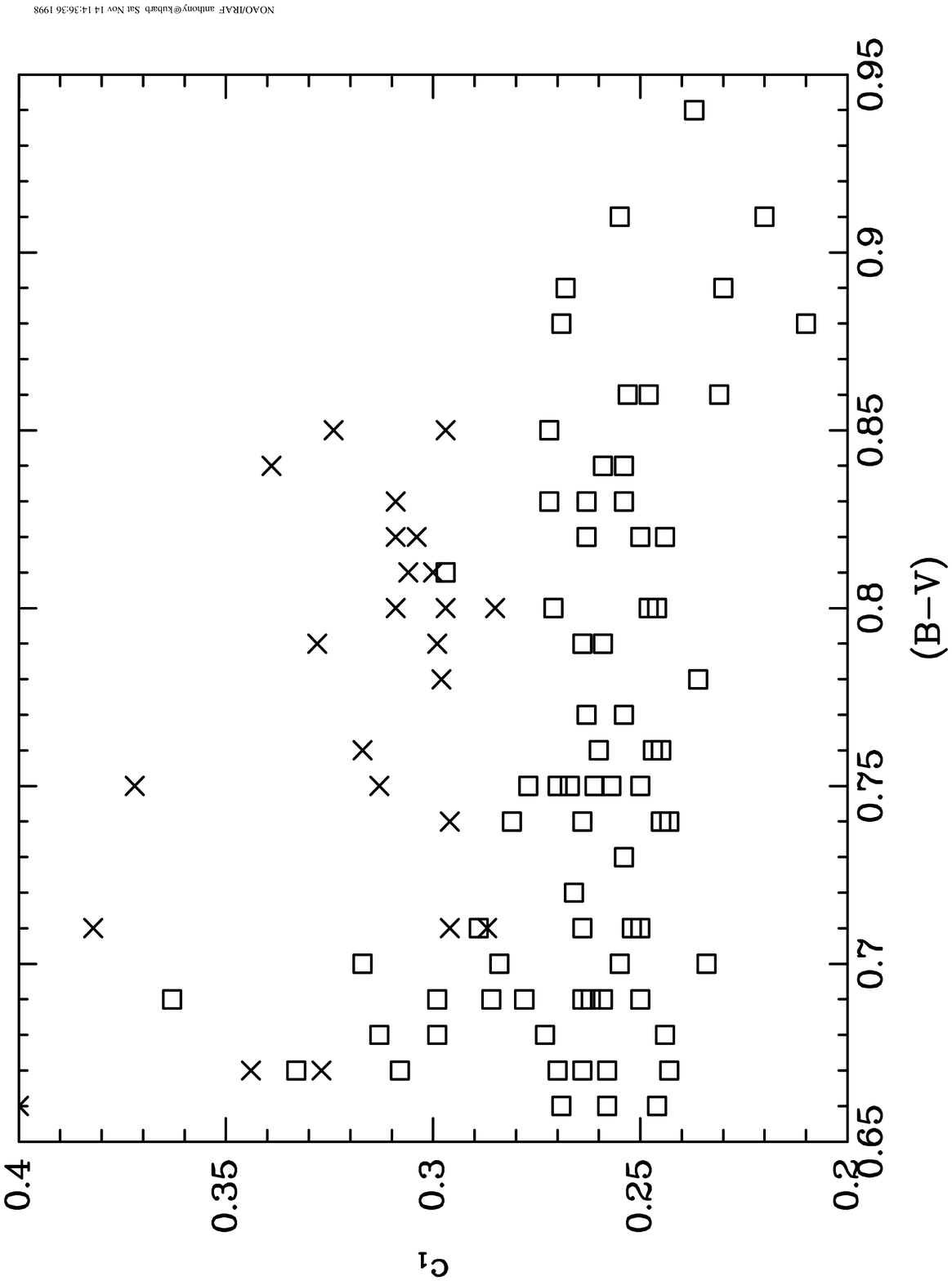]{The observed trend between $c_1$ and $(B-V)$ for the stars in
Fig. 5. Crosses are stars tagged as subgiants from Fig. 5, while squares
are stars on the unevolved main sequence and probable binaries.\label{fi6}}
\figcaption[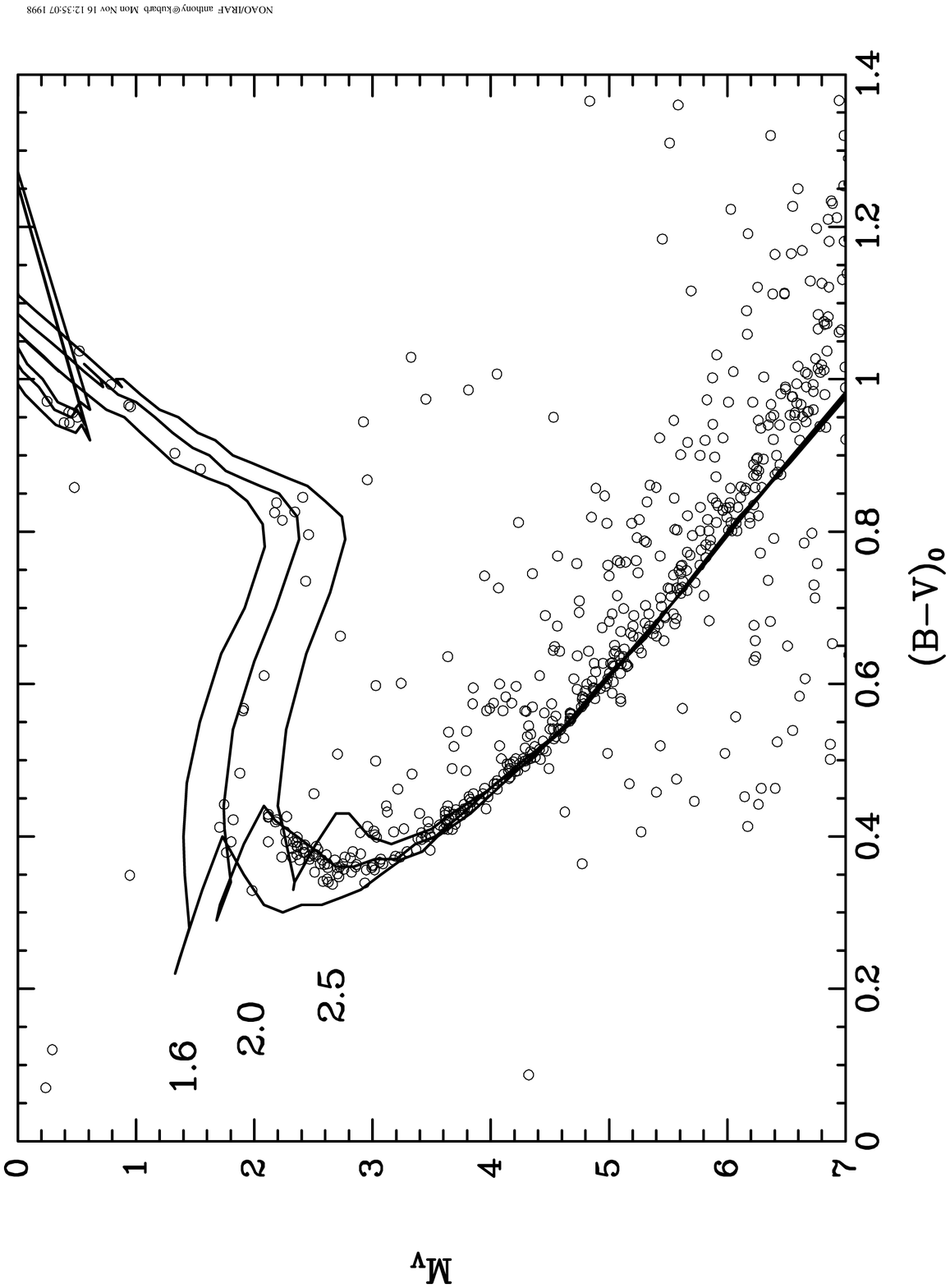]{Comparison of the CMD of NGC 2420 with the isochrones of
BE for [Fe/H] = --0.4, adjusted to [Fe/H] = --0.3. Adopted cluster parameters
are $E(B-V)$ = 0.04 and $(m-M)$ = 12.15. Isochrones are identified by their
age in Gyr.\label{fi7}}
\figcaption[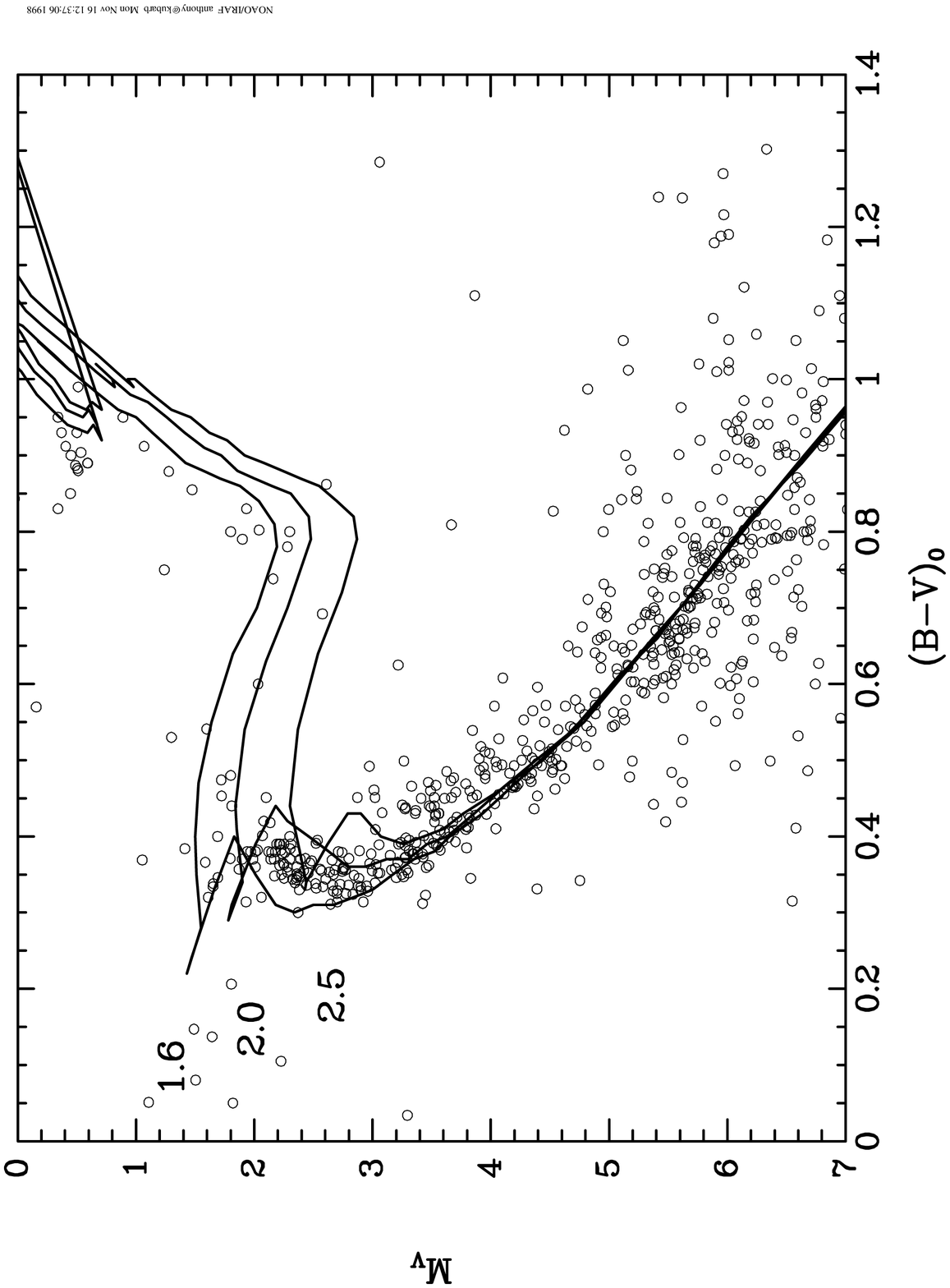]{Comparison of the CMD of NGC 2506 with the isochrones of BE for
[Fe/H] = --0.4. Adopted cluster parameters are $E(B-V)$ = 0.04 and $(m-M)$ =
12.70. Isochrones are identified by their ages in Gyr.\label{fi8}}
\figcaption[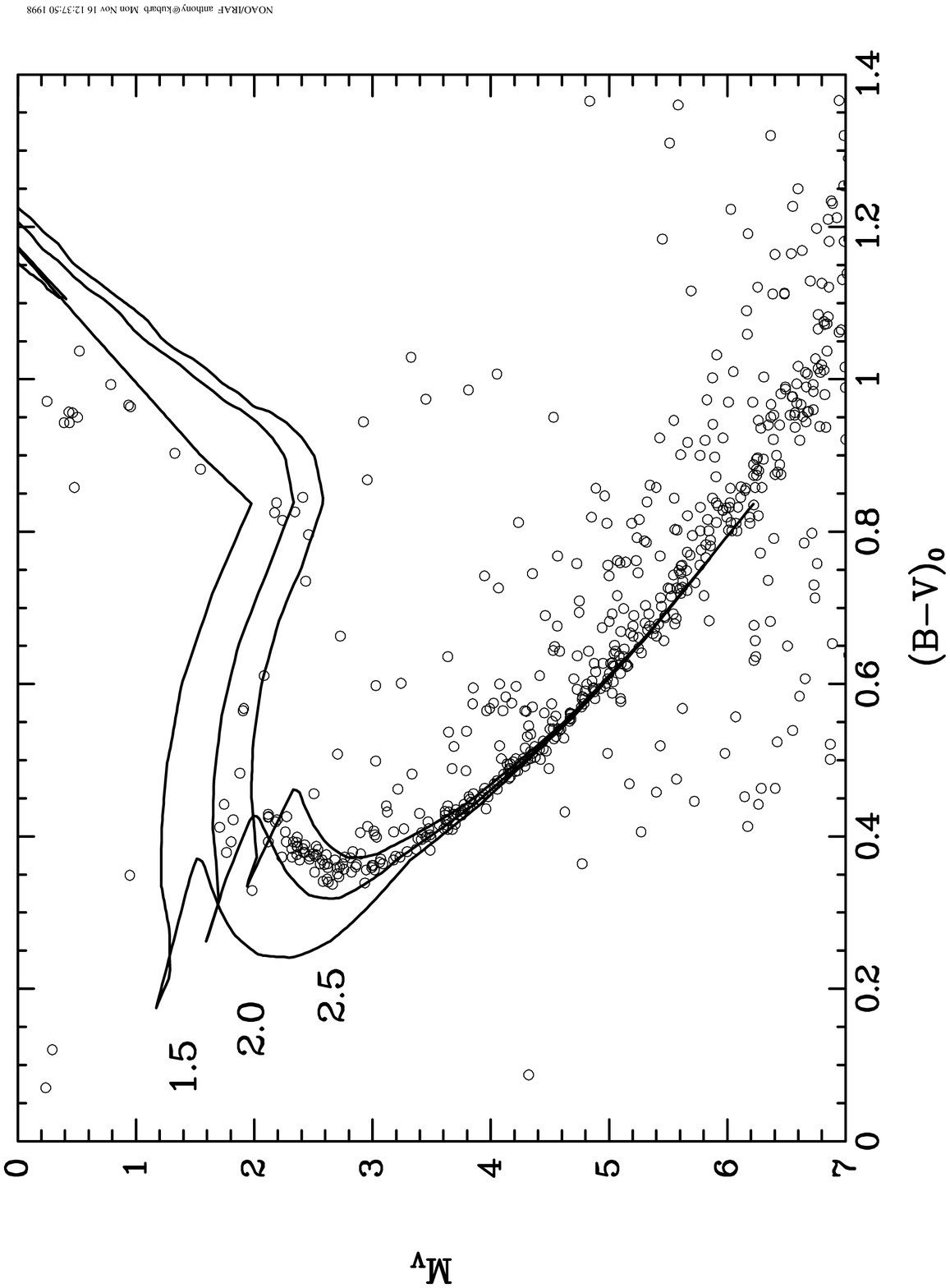]{Same as Fig. 7 for the isochrones of GE. \label{fi9}}
\figcaption[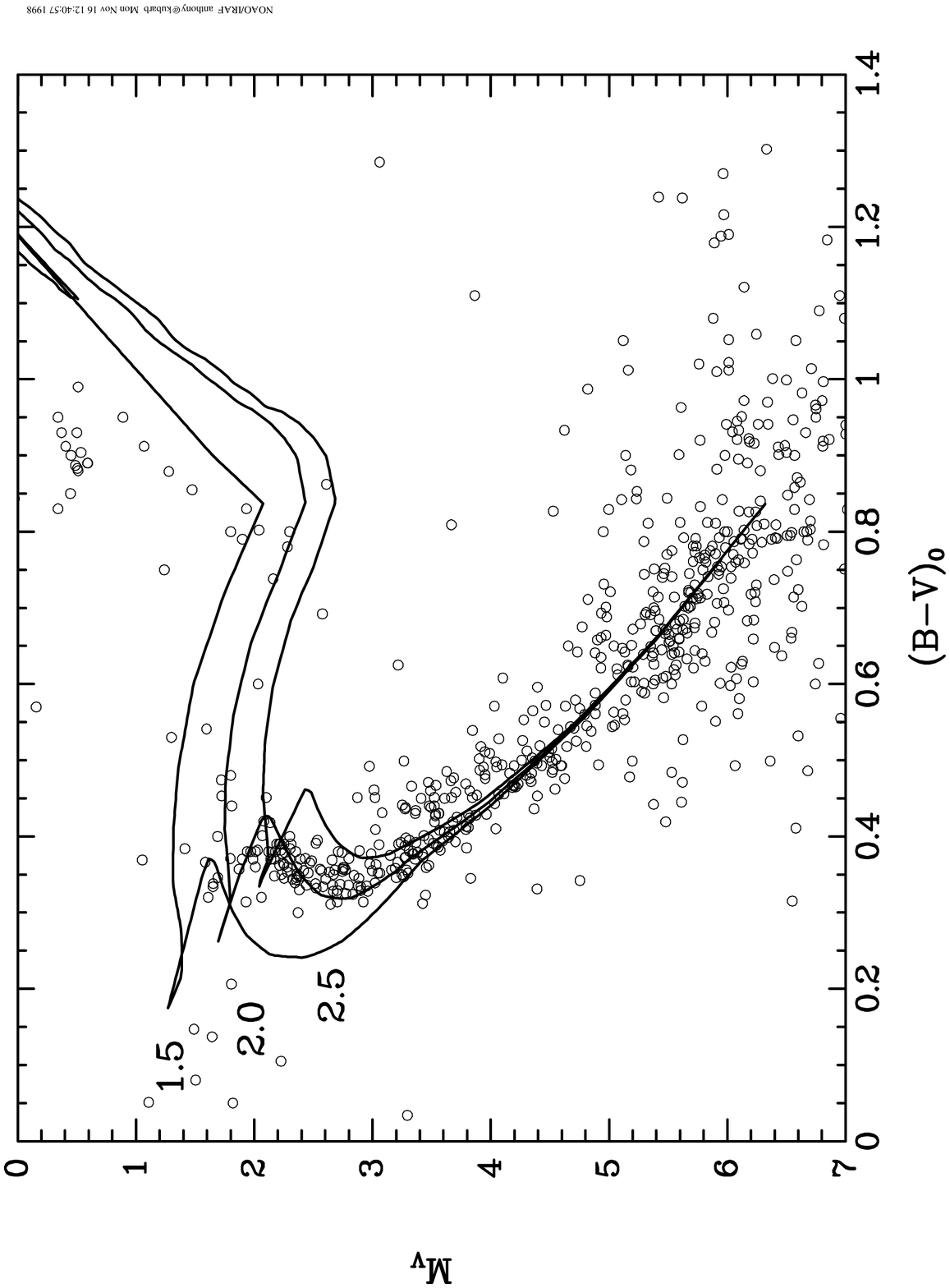]{Same as Fig. 8 for the isochrones of GE. \label{fi10}}
\figcaption[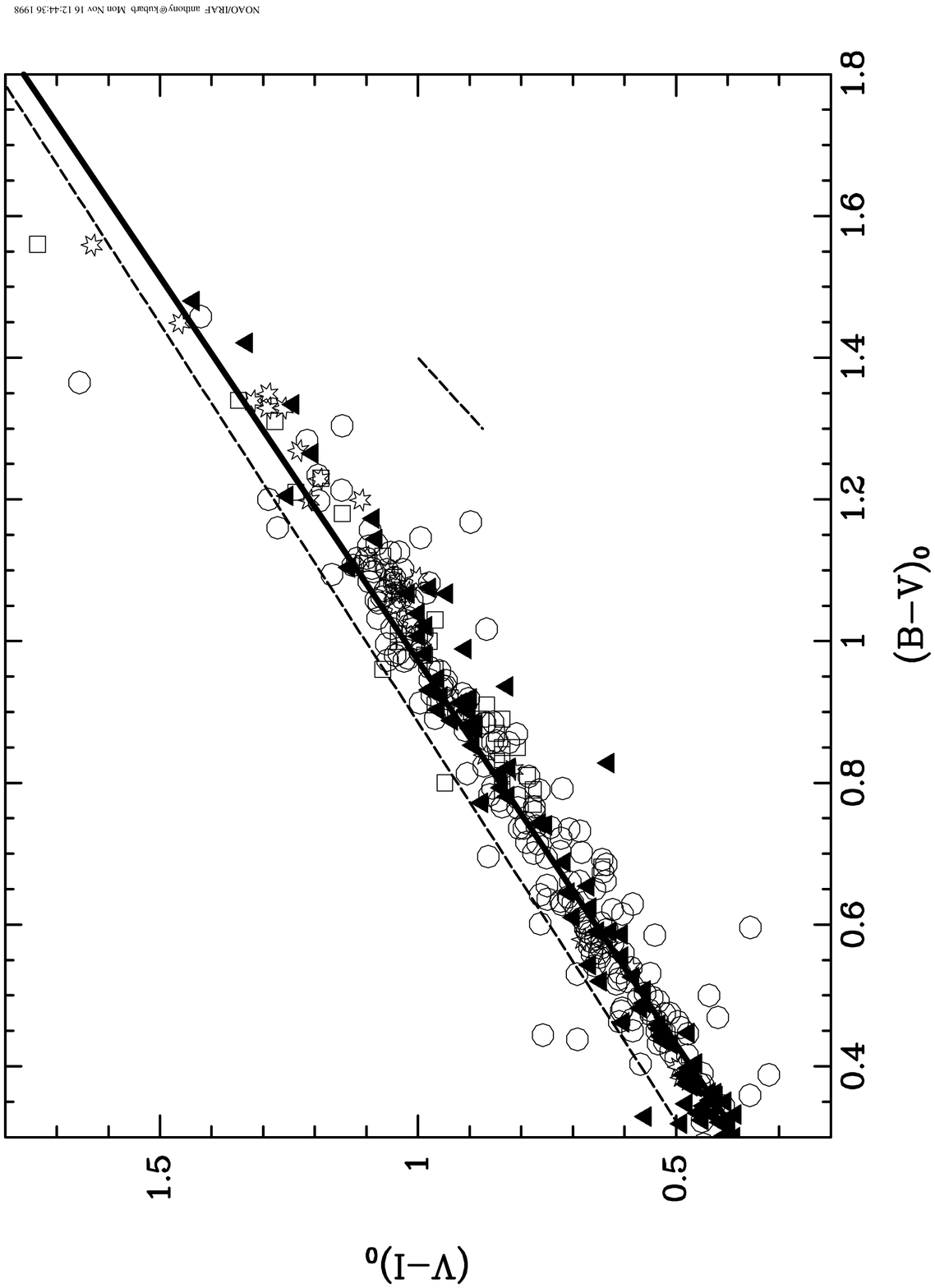]{Correlation between $(V-I)_0$ and $(B-V)_0$ for the
giants in NGC 2204 (filled triangles), Mel 66 (open squares), 
Be 39 (open circles), and M67 (stars). \label{fi11}}
\figcaption[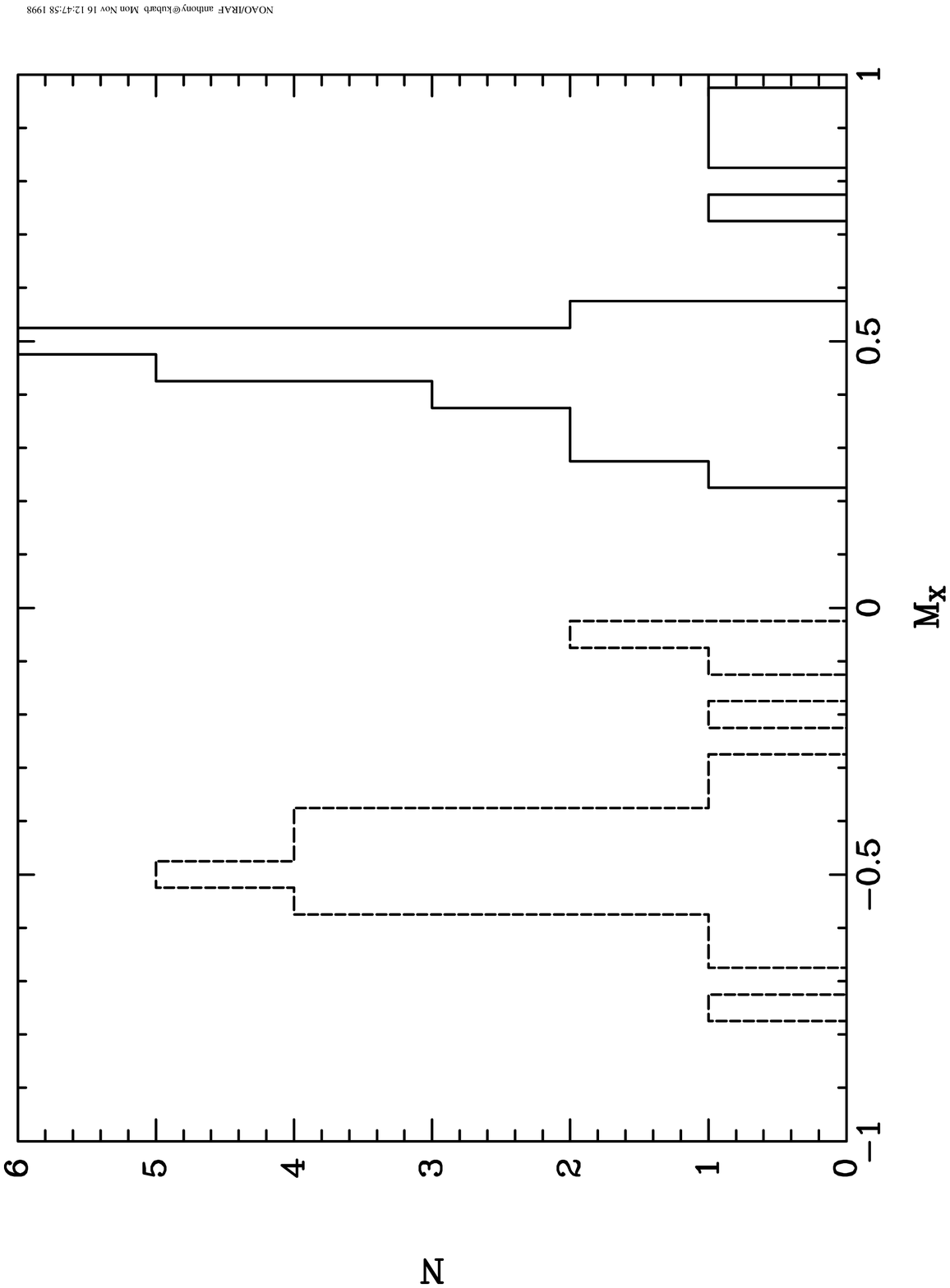]{Distribution of combined RGC sample for NGC 2506 and NGC 2420
as a function of $M_I$ (dashed curve) and $M_V$ (solid curve). \label{fi12}}

\begin{table}
\dummytable\label{tab1}
\end{table}
\end{document}